# The Relevance of Non-axiality and Low-lying Excited States for Slow Magnetic Relaxation in Pentagonal-bipyramidal Erbium(III) Complexes Probed by High-frequency EPR

Jan Arneth,*[a] Lena Spillecke,[a] Changyun Koo,[a],[c] Tamara A. Bazhenova,[b] Eduard B. Yagubskii,[b] and Rüdiger Klingeler*[a]

[a] J. Arneth, Dr. L. Spillecke, Dr. C. Koo, Prof. Dr. R. Klingeler
Kirchhoff Institute for Physics
Heidelberg University
INF 227, D-69120 Heidelberg, Germany
E-mail: jan.arneth@kip.uni-heidelberg.de, ruediger.klingeler@kip.uni-heidelberg.de

[b] Dr. T. A. Bazhenova, Prof. Dr. E. B. Yagubskii
Federal Research Center of Problems of Chemical Physics and Medicinal Chemistry
Russian Academy of Sciences
Ac. Semenov Avenue 1, 142432 Chernogolovka, Moscow region, Russian Federation

[c] New affiliation: Dr. C. Koo
Department of Physics
Sungkyunkwan University
Suwon 16419, Republic of Korea

**Abstract:** High-frequency/high-field electron paramagnetic resonance studies on a series of seven-coordinate pentagonal-bipyramidal (PBP) erbium(III) complexes Er(DAPMBH/H$_2$DAPS)X (H$_2$DAPMBH = 2,6-diacetylpyridine bis-4-methoxy benzoylhydrazone, H$_4$DAPS = 2,6-diacetylpyridine bis-(salicylhydrazone)) demonstrate the effects of different apical ligands (X = (H$_2$O)Cl **(1)**, (CH$_3$OH)N$_3$ **(2)**, Cl$_2$ **(3)**) on the local magnetic anisotropy of the central Er$^{III}$ ions. In particular, we report direct experimental determination of the effective $g$-values and zero field splittings of the energetically low-lying Kramers doublets. Our quantitative determination of the magnetic anisotropy highlights the relevance of an axial $g$-tensor for SMM behaviour and suggests that fast magnetic relaxation is mainly driven by a thermally assisted quantum tunnelling process via low-lying excited states.

## Introduction

The discovery of magnetic bistability, i.e., the ability to retain magnetisation for long times in the absence of an external magnetic field, at liquid helium temperatures in the famous dodecanuclear Mn$_{12}$ac-molecule[1] was followed by an intense search for further single molecular magnets (SMM) with ever higher blocking temperatures ($T_B$).[2-5] Due to their molecular nature, SMMs exhibit unique advantages over bulk magnetic materials, and a wide variety of possible applications in high-density data storage and processing and in quantum computing have been put forward.[6-8] In the early years of molecular magnetism, research was mainly focused on designing polynuclear 3$d$ coordination clusters with high total spins and large anisotropy barriers.[9-11] Quickly, also the idea of implementing lanthanide ions for synthesizing SMMs emerged since intrinsically strong spin-orbit coupling and large unquenched orbital momentum of 4$f$ moments lead to considerably enhanced single-ion anisotropy and, hence, large effective energy barriers ($U_{eff}$).[12-13] Furthermore, using rare-earth ions opens the possibility of constructing SMMs with only a single magnetic center, termed single ion magnets (SIM). However, it was not until about almost 10 years after the discovery of the first SMM that slow magnetic relaxation was demonstrated in the double-decker [LnPc$_2$]$^-$ (Ln = Tb, Dy) compounds.[14] From there on, an increasing amount of effort has been put into the design of lanthanide-based SIMs, and record values for effective energy barriers of $U_{eff}$ = 1815 K and blocking temperatures as high as $T_B$ = 80 K have been reached in mononuclear dysprosium complexes.[15-16]

In particular, it has become clear that engineering magnetic anisotropy is one of the most promising ways to design high-performance lanthanide-based SMMs.[17-18] Placing the magnetic center in a suitable coordination environment, i.e., controlling the coordination geometry and the local symmetry around the 4$f$ moment, results in strong uniaxial magnetic anisotropy and yields quenching of the quantum tunnelling of magnetisation (QTM). In this context, mononuclear Dy$^{III}$ complexes have been shown to exhibit substantial SMM characteristics when the magnetic ion is coordinated in a pentagonal-bipyramidal (PBP) crystal field.[15, 19-20] The pseudo-$D_{5h}$ symmetry yields a large crystal field splitting of the lowest $^6H_{15/2}$ multiplet and an almost strict axiality of the crystal field potential, rendering the $m_J$ = 15/2 Kramers doublet (KD) with an Ising-like effective $g$-tensor ($g_{xx}$ = $g_{yy}$ ≈ 0, $g_{zz}$ ≈ 20) the magnetic ground state.[15, 21-24]

Despite the quite active and successful work on PBP Dy$^{III}$ complexes, research on equally seven-coordinated Er$^{III}$-containing compounds, which also exhibit a high total angular momentum of $J$ = 15/2 in the ground state, is hardly reported in the literature.[25-27] Generally, only few Er$^{III}$-based SMM are known, which arises from the need for equatorial ligand fields to effectively enhance the magnetic single ion anisotropy (SIA) of the predominantly prolate electron distribution, and quantitative analysis of the SIA is rarely performed.[28-30] Recently, Bazhenova et al. synthesized and studied a series of pentagonal-bipyramidal erbium complexes with acyclic chelating N$_3$O$_2$ Schiff-base ligands in the equatorial plane and varying apical ligands.[31] Within this series field-induced SMM behaviour ($U_{eff}$ ≈ 25 K) is observed when the central Er$^{III}$ ion is coordinated by one neutral and one charged axial ligand, while the complex (Et$_3$NH)[Er(H$_2$DAPS)Cl$_2$] (H$_4$DAPS = 2,6-diacetylpyridine bis-(salicylhydrazone)) **(3)** with two negatively charged apical chlorine ligands exhibits fast magnetic relaxation. Numerical crystal field analysis of the dc magnetisation and *ab-initio* calculations suggest that slow relaxation of the magnetisation occurs due to the strong axiality of the crystal field, as demonstrated by the dominance of a positive $B_{40}$ crystal field parameter. For complex **(3)** the different theoretical approaches yield significantly contrary results regarding the axiality of the ground state *g*-tensor and the size of the zero field splittings to the excited KDs.

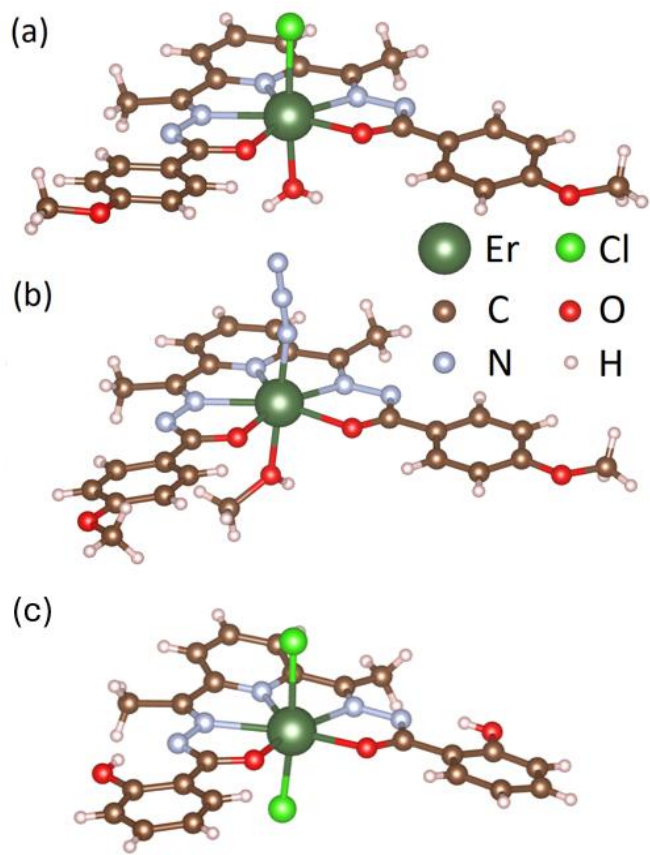

**Figure 1.** Molecular structure of the investigated pentagonal-bipyramidal Er$^{III}$-complexes **(1)** (a), **(2)** (b), and **(3)** (c). Figures were generated by VESTA[36] using the crystallographic information from Ref. 31.

To shed light on these discrepancies we performed high-frequency/high-field electron paramagnetic resonance (HF-EPR) spectroscopy measurements on **(3)** in an earlier study.[32] Our experimental data validate the findings of numerical crystal field analysis, that are, the presence of energetically low-lying KDs ($\Delta_{1\rightarrow 2}$ = 13.9 K and $\Delta_{1\rightarrow 3}$ = 22.1 K) and considerable non-axiality of $g_{eff}$ ($g_{eff,x}$ = 2.6, $g_{eff,y}$ = 3.2 and $g_{eff,z}$ = 12.5). In the present work we extend this study by reporting HF-EPR spectroscopy data on the equally pentagonal-bipyramidal coordinated erbium complexes Er(DAPMBH)X (H$_2$DAPMBH = 2,6-diacetylpyridine bis-4-methoxy benzoylhydrazone and X = (H$_2$O)Cl **(1)**, (CH$_3$OH)N$_3$ **(2)**) from the same series which allows us to elucidate in detail the effect of variation of the apical ligands on the low-lying electronic states and, hence, on the electronic and magnetic properties in this series. In addition, a direct comparison of our experimental HF-EPR results with numerical studies allows us to validate numerical calculations and to critically assess the correctness of *ab-initio* predictions as done recently by us for various lanthanide complexes.[33-35]

## Experimental Results

The complexes investigated in this work were synthesized and characterized as reported in Ref. 31, and their molecular structures are displayed in Fig 1. Analogously to the previously reported complex **(3)**[32] [Fig. 1(c)], the Er$^{III}$ ions in both of the here studied compounds **(1)** and **(2)** are surrounded by a pentadentate ligand plane consisting of two oxygen and three nitrogen ions in the first coordination sphere. The key difference between the distinct molecules is, hence, given by the axial ligands: While the apical coordination sites in **(1)** and **(2)** are occupied by one charged group (Cl$^-$, N$_3^-$) and one neutral molecule (H$_2$O, CH$_3$OH), **(3)** hosts a charged chloride ion at both positions leading to a net negative charge of the complex.

To investigate the magnetic properties and decipher the origin of different relaxation dynamics, HF-EPR measurements were performed on Er$^{III}$ complexes **(1)** and **(2)**. The absorption spectra obtained from oriented loose powder samples of both complexes at *T* = 2 K are shown in the background of Fig. 2(a) and (b). For comparison, Fig. 2(c) depicts the low-temperature loose powder HF-EPR spectra of the charged complex **(3)** with similar equatorial ligands but two chloride ions in the apical positions as reported in Ref. 32. As can be seen, all spectra exhibit distinct Lorentzian-shaped resonance features in the whole measured frequency range 0 – 900 GHz. The observed homogenous line broadening arises from the pseudo-single-crystal nature of the loose powder spectra, in which only a single molecular direction is probed. While the main absorption peak constitutes the only feature visible in complex **(1)** at the lowest temperature, complex **(2)** and **(3)** exhibit two additional resonance features above 800 GHz and 300 GHz, respectively. The extracted resonance field positions at several fixed frequencies are summarized in Fig. 2. As pointed out by the solid black lines, the resonance features form clear branches, which are labelled as (i) for **(1)**, α to γ for **(2)**, and R1 to R3 for **(3)**. All observed branches show a linear field dependence, except for the field region around 5 T in **(2)** and the crossing regime of R1 and R2 in **(3)**. The latter indicates avoided crossing behaviour, i.e., mixing of different energy states. While the branches (i), α and R1 are gapless within the resolution of our HF-EPR experiment, extrapolation of the resonance branches appearing at higher frequencies in **(2)** and **(3)** yields zero-field splittings (ZFS) of 760 GHz and 290 GHz, respectively. The effect of increasing temperature on the HF-EPR spectra of **(1)** at an exemplary fixed frequency of $f = 593.9$ GHz is shown in Fig. 3(a). As the temperature rises, the pronounced absorption peak (black square) gradually decreases in intensity until it is almost indiscernible at $T = 60$ K. Hence, the resonance branch (i) is associated with a ground state transition. Additionally, our data reveal another resonance feature (red circle) evolving at elevated temperatures around $T \approx 8$ K, thereby implying the presence of energetically higher-lying spin states that are thermally populated. Similar to the ground-state transition, the resonance frequency of the excited-state feature also follows a linear branch, labelled as (ii) in the resonance frequency vs. magnetic field diagram (Fig. 3(b)), but exhibits a ZFS of 825 GHz. The resonance branches obtained on a loose powder sample of **(1)** at various temperatures and displayed in Fig. 3(b) provide direct information on the low-lying excited energy levels. As mentioned above, the Curie-like temperature dependence and the absence of ZFS in (i) clearly implies that it is associated with a transition within the lowest Kramers doublet (KD1). In contrast, a finite resonance frequency in zero magnetic field, along with its appearance at higher temperatures, renders (ii) a transition between two KDs starting from an excited energy state. From the slope of the branches at small magnetic fields, we read off effective g-values $g_{eff}$ of 12.0(3) and 9.4(5) for branches (i) and (ii), respectively. Applying a phenomenological $S = 1/2$ pseudospin approximation for each individual KD of the $^4I_{15/2}$ ground state multiplet (cf. Ref. 32) yields the energy level diagram shown in Fig. 3(c). Since the experimentally accessible energy range does not allow probing higher-lying KDs, only the two energetically lowest KDs are shown in Fig. 3(c), while further energy states are omitted. Note here that the construction of this energy level diagram is solely based on the effective g-values and ZFS observed for the branches (i) and (ii). Our analysis yields effective *g*-values of 12.0(3) and 6.8(4) for KD1 and KD2, respectively.

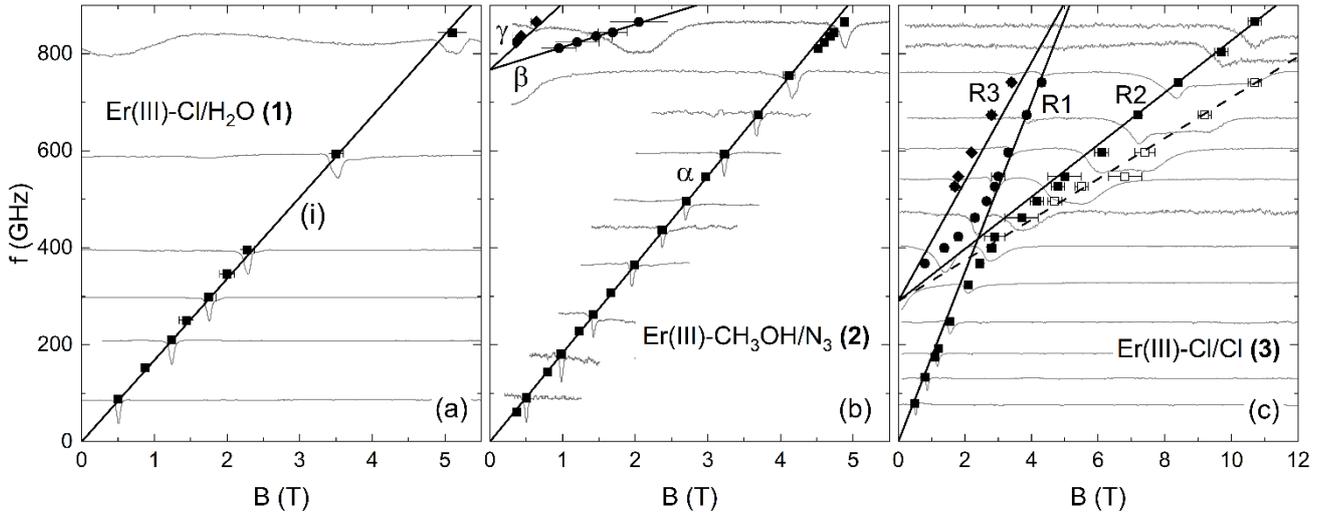

**Figure 2.** Resonance frequency versus magnetic field diagram for a loose powder sample of complex **(1)** (a), **(2)** (b), and **(3)**[32] (c) obtained at $T = 2$ K. Solid black lines depict simulated ground state transitions using a $S = 1/2$ pseudo-spin approximation for each KD (see the text). Grey lines show the measured HF-EPR spectra, which are shifted along the ordinate for better comparison with the corresponding resonance feature.

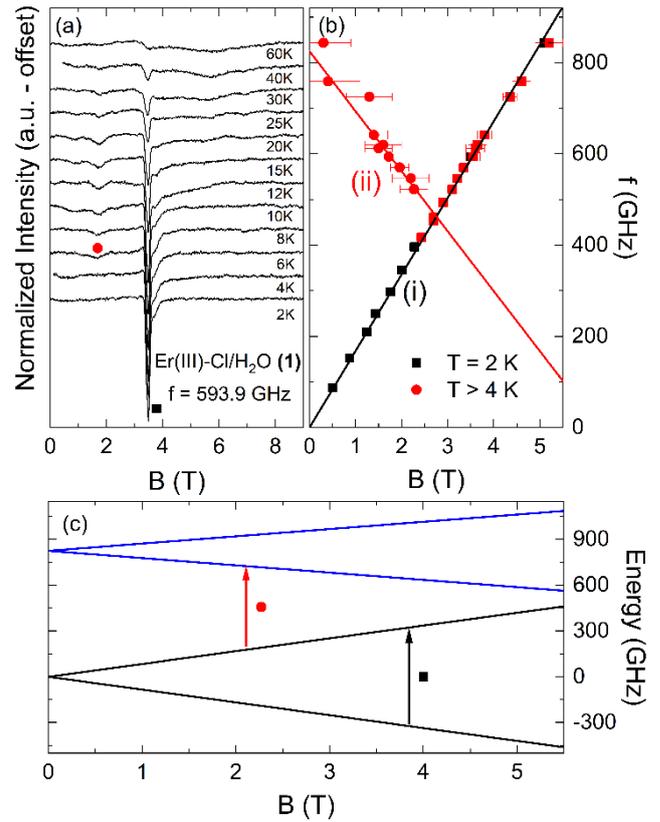

**Figure 3.** Loose powder HF-EPR spectra of complex **(1)** measured at selected fixed temperatures at $f = 593.9$ GHz (a), the corresponding frequency-field-diagram (b), and simulated energy levels of the two energetically lowest KDs (c). Solid lines in (b) and (c) depict simulations using a $S = 1/2$ pseudospin approximation (see the text). The arrows mark transitions between the energy levels corresponding to the observed resonance branches. For the HF-EPR spectra at elevated temperatures obtained at various frequencies see the Supplemental Material (SM).

Due to the alignment of the loose powder sample with the external magnetic field, the obtained spectra correspond to a measurement along the easy-anisotropy axis, and thus, only the corresponding component of the *g*-tensor can be quantified from these measurements. In order to obtain the complete *g*-tensor, and hence to further elucidate magnetic anisotropy, fixed powder HF-EPR measurements on complex **(1)** have been performed. Exemplary fixed powder spectra obtained at $f$ = 252.3 GHz and $f$ = 168.4 GHz are displayed in Fig. 4. Two resonance features can be identified in the spectra. By comparing the fixed powder spectra to the loose powder spectra in Fig. 2(a), the feature at lower magnetic fields can be assigned to the easy axis orientation, here denoted as the *z*-direction, with $g_{eff,z}$ = 12.0(1) (see Fig. S3 for a direct comparison of fixed and loose powder spectra). The *g*-values corresponding to the other crystallographic directions ($g_{eff,x}$ and $g_{eff,y}$) are implied to be considerably smaller. In particular, the almost equal intensity of the resonance features indicates that only two directions contribute to the measured spectra while the third *g*-value component is most likely too small to yield a corresponding feature in the accessible field range, i.e., $g_{eff,z} > g_{eff,y} \gg g_{eff,x}$. To quantify the *g*-tensor components the obtained HF-EPR fixed powder spectra were fit assuming an $S$ = 1/2 pseudospin approximation with a strongly anisotropic *g*-tensor (solid red lines in Fig. 4). The best agreement with the experimental data was achieved by using $g_{eff,z}$ = 12.0(1), $g_{eff,y}$ = 2.6(5) and $g_{eff,x}$ < 0.5(5), where $g_{eff,x}$ is fixed to the highest possible value which still reasonably reproduces the measured spectra while the other two components are varied.

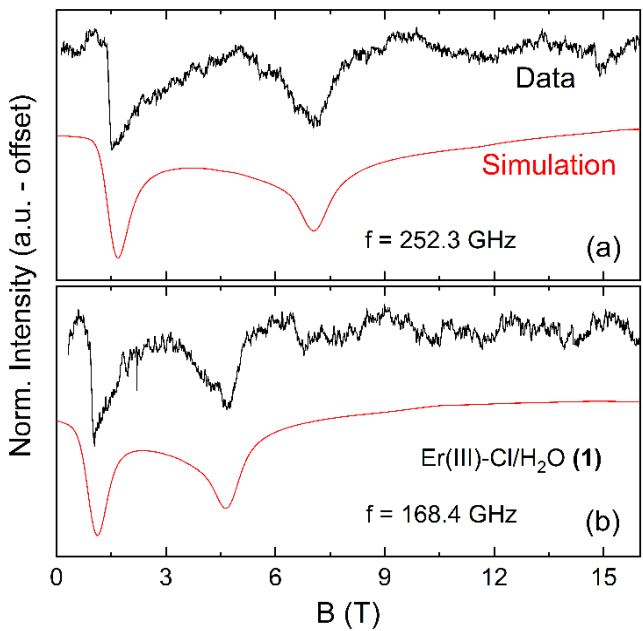

**Figure 4.** HF-EPR spectra (black lines) of a fixed powder sample of complex **(1)** measured at $f$ = 252.3 GHz (a) and $f$ = 168.4 GHz (b) at $T$ = 2 K. Red lines depict simulated spectra using a $S$ = 1/2 pseudo-spin approximation with a strongly anisotropic *g*-tensor as described in the text.

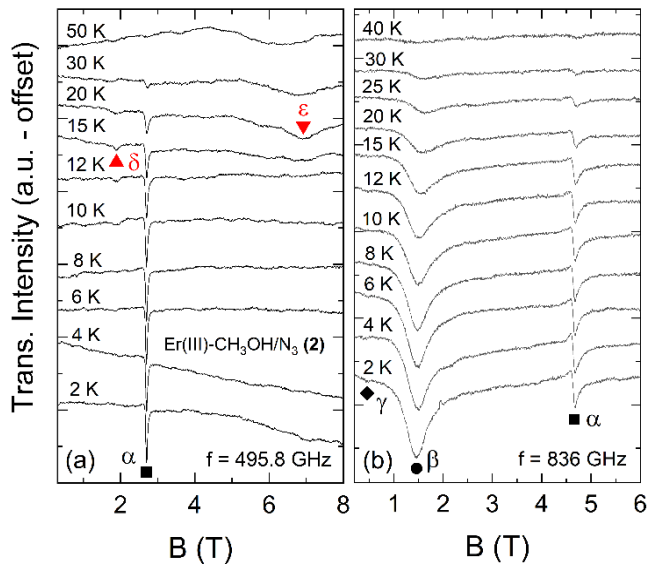

**Figure 5.** Temperature dependence of the loose powder HF-EPR spectra of complex **(2)** measured at fixed frequencies of $f$ = 495.8 GHz (a) and $f$ = 863 GHz (b). The symbols mark resonance features appearing at T = 2 K (black) and at higher temperatures (red).

Continuing the investigation of complex **(2)**, Fig. 5 displays the temperature dependence of the loose powder spectra measured at frequencies of $f$ = 495.8 GHz (a) and $f$ = 836.0 GHz (b). Similar to the observation in **(1)**, the spectral weight of all resonances that can be identified at $T$ = 2 K, i.e., α, β, and γ, decreases with rising temperature, which suggests that they are associated with transitions from the magnetic ground state. Further, the appearance of two additional resonance features upon heating (red triangles), here labelled as δ and ε, implies transitions from excited energy states which are thermally populated.

A summary of all resonance positions observed at different temperatures in a loose powder sample of **(2)** is given in Fig. 6(a). In total, we observe five resonance branches, all of which follow a linear field dependence up to a magnetic field of approximately 5 T. Above this field, an avoided crossing behaviour of branch ε confirms the mixing of energy states, as already implied by the weak right-bending of branch α. The absence of a finite excitation gap at zero magnetic field allows for the assignment of resonance branch α to a transition within the energetically lowest Kramers doublet (KD1). In contrast, branches β - ε are gapped with a ZFS of Δ = 767 GHz and, hence, associated with transitions from the lowest to the first excited KD (KD2). As depicted in Fig. 6(b), the observed resonance positions can be rationalized again in a phenomenological pseudospin $S$ = 1/2 approximation for the two lowest-lying KDs, where the arrows assign the resonance branches to the transitions from the ground state (black) or the first excited state (red). By fitting the ground state branches α – γ, effective *g*-factors of $g_{eff}$ = 13.1(1) and 6.2(3) can be calculated for KD1 and KD2, respectively. The consistency of the used model is demonstrated by a very good agreement with the observed transitions δ and ε as indicated by the simulated branches (solid lines) in Fig. 6(a).

The anisotropic nature of complex **(2)** can almost directly be seen from the fixed powder HF-EPR spectra, one of which is exemplarily shown in Fig. 7(a). Similar to the findings in complex **(3)**, the fixed powder spectra of **(2)** exhibit the typical shape for an XXZ-type $g$-tensor anisotropy in the case of a randomly oriented powder sample, and two distinct resonance features can be identified.[32, 37] The resonance positions of both observed features at different measurement frequencies form two clear resonance branches, as displayed in Fig. 7(b). By comparison to Fig. 6(a), the linear branch appearing at lower magnetic fields (solid squares) can be straightforwardly assigned to the resonances arising from the easy axis direction as observed in the loose powder measurement (see Fig. S6 for a direct comparison of fixed and loose powder spectra), while the stronger resonance feature at higher fields (open squares) corresponds to another crystallographic direction within the molecular complex. Since the external magnetic field rotates the Er$^{III}$ moments out of their zero-field easy axis orientation, the branch associated with the transversal molecular direction exhibits slight left-bending to higher magnetic fields, which complicates the determination of an effective $g$-value. In order to at least estimate $g_{eff,x/y}$, the HF-EPR spectra obtained at small frequencies, i.e., in a regime where the corresponding branch is in good approximation linear, were simulated analogous to the procedure used for **(1)** and illustrated in Fig. 4(a). The best fit to the data is depicted as a red solid line in Fig. 7(a). From the intensity distribution and the positions of the resonance features, an XXZ-type $g$-tensor with principal values of $g_{eff,z} = 13.1(2)$, $g_{eff,y} = 1.5(1)$, and $g_{eff,x} = 1.5(3)$ can be derived.

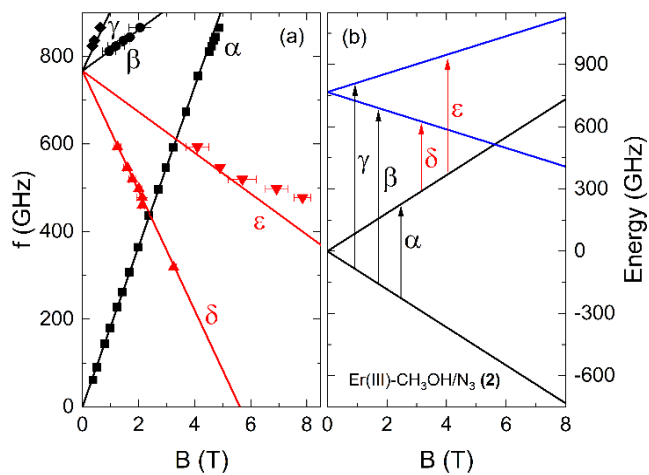

**Figure 6.** Frequency versus magnetic field diagram of complex **(2)** obtained from loose powder HF-EPR measurements at different temperatures (a) and simulated energy levels including the two energetically lowest KDs (b). Solid lines depict simulations using a $S = 1/2$ pseudospin approximation for each KD as described in the text. The arrows mark transitions between the energy levels corresponding to the observed resonance branches. For the HF-EPR spectra at elevated temperatures obtained at various frequencies see the Supplemental Material (SM).

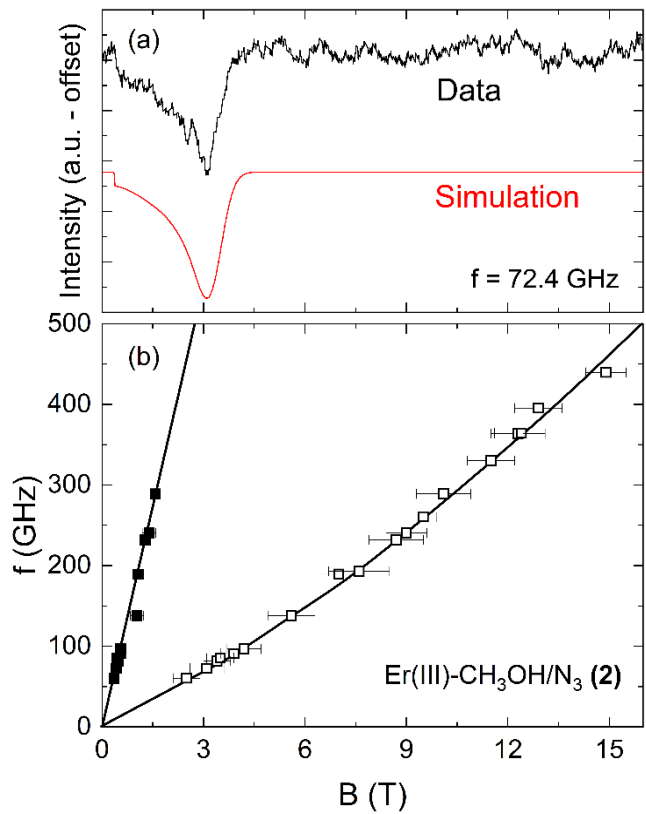

**Figure 7.** (a) Fixed powder HF-EPR spectrum (black) of complex **(2)** measured at $f = 72.4$ GHz and $T = 2$ K and simulation assuming a $S = 1/2$ pseudospin approximation (red) as described in the text. (b) Frequency versus magnetic field diagram obtained at $T = 2$ K including the loose (closed) and fixed (open) powder resonance features. All lines in (b) are a guide to the eye. For the HF-EPR spectra at higher frequencies see the Supplemental Material.

## Discussion

Tab. 1 summarizes the anisotropic $g$-values and crystal field splittings of complexes **(1)** - **(3)** as experimentally determined by simulating the HF-EPR spectra in an effective $S = 1/2$ pseudospin model for each KD. In addition, we cite the results of numerical studies based on crystal field (CF) analysis of magnetic data and *ab-initio* calculations, respectively, from Ref. 31. For all three complexes under study, the ground-state $g$-tensor shows a pronounced uniaxial behaviour ($g_{eff,z} \gg g_{eff,x}, g_{eff,y}$) which suggests an Ising-like character in the low-temperature magnetic behaviour of the Er$^{III}$ ions. However, the presence of finite $g_{eff,x}$ and $g_{eff,y}$ values directly implies considerable contributions from non-axial crystal field parameters as predicted by numerical studies.[31] For **(2)**, the Ising character is most pronounced and $g_{eff,x} \approx g_{eff,y}$ is found within the resolution of the experiment. Concomitantly, **(2)** features the largest anisotropy barrier of $U_{eff} = 28$ K among the studied complexes.[31] In contrast, the $g$-tensors of **(1)** and **(3)** exhibit larger transversal components with $g_{eff,x} \neq g_{eff,y}$, which probably mainly arises from the distorted equatorial $N_3O_2$ pentagon (Fig. 1).[31-32] The presence of non-axial crystal-field parameters leads to a mixing of the $\pm m$ states and, hence, favours a fast QTM in the ground state KD.[38]

In order to gain further insights into the crystal field effects on the spin manifold, we turn back to the pseudo-single-crystalline loose powder spectra: The large effective *g*-values observed in particular for the ground state KDs in all studied compounds imply that the observed resonance features arise from strongly forbidden transitions with $\Delta m > 1$. By assuming that the theoretical free-ion Landé-factor $g_L = 1.2$ for Er$^{III}$ is in good approximation still valid in the presence of the pentagonal-bipyramidal ligand coordination, the effective magnetic quantum number of the corresponding KDs can be determined via $m_{eff} = \Delta m/2 = g_{eff,z}/2g_L$. The so calculated $m_{eff}$ are summarized in Tab. 2. The half-integer $m_{eff}$ values found for KD1 and KD2 in complex **(2)** directly indicate an almost pure $m = 11/2$ and $m = 5/2$ composition of the magnetic ground state and the first excited state, respectively. While the ground state $m_{eff}$ in **(3)** is also close to a half-integer value within its error bars, our analysis implies a clear mixing of states in complex **(1)**. The observed trend is in qualitative agreement with the finding of **(2)** exhibiting the smallest transversal anisotropy (Tab. 1), which is directly responsible for the occurrence of state mixing. Further, it becomes evident that $m = 11/2$ is the main contribution to the spin wavefunction in the magnetic ground state of all studied compounds, which confirms the dominance of a positive $B_{40}$ crystal field parameter as suggested by crystal field analysis.[31]

**Table 1.** Effective *g*-values and zero field splittings of the up to three energetically lowest Kramers doublets in the complexes **(1)** – **(3)** as obtained from our analysis of the HF-EPR measurements. For comparison, the corresponding values determined by means of numerical studies based on crystal field analysis (CF) of the dc magnetisation and *ab-initio* calculations from Ref. 31 are shown.

|  | Er$^{III}$-Cl/H$_2$O **(1)** slow relaxation ($U_{eff}$ = 23 K)[31] | | | Er$^{III}$-CH$_3$OH/N$_3$ **(2)** slow relaxation ($U_{eff}$ = 28 K)[31] | | | Er$^{III}$-Cl/Cl **(3)** SMM-silent[31] | | | |
|---|---|---|---|---|---|---|---|---|---|---|
|  | $g_{eff,x}$ | $g_{eff,y}$ | $g_{eff,z}$ | $g_{eff,x}$ | $g_{eff,y}$ | $g_{eff,z}$ | $g_{eff,x}$ | $g_{eff,y}$ | $g_{eff,z}$ |  |
| KD1 | 12.0(3) | 2.5(3) | < 0.5(5) | 13.1(2) | 1.5(1) | 1.5(3) | 12.5(4) | 3.2(3) | 2.6(2) | this work |
|  | 13.44/10.9 | 2.10/5.17 | 0.53/0.03 | 14.20/11.51 | 2.0/4.7 | 0.1/0.46 | 12.37/15.04 | 4.88/0.55 | 2.07/0.44 | CF/*ab-initio*[31] |
| KD2 | 6.8(5) | - | - | 6.2(3) | - | - | 4.9(2) | - | - | this work |
|  | 9.68/6.16 | 5.09/2.64 | 1.96/0.35 | 8.92/5.05 | 5.24/1.80 | 2.75/0.39 | 7.75/4.14 | 6.34/6.73 | 2.70/7.31 | CF/*ab-initio*[31] |
| KD3 | - | - | - | - | - | - | 3.2(2) | - | - | this work |
| $\Delta_{1\to2}$ | 825(13) GHz | | | 767(8) GHz | | | 290(11) GHz | | | this work |
|  | 873/992 GHz | | | 655/786 GHz | | | 270/785 GHz | | | CF/*ab-initio*[31] |
| $\Delta_{1\to3}$ | ≫ 1 THz | | | ≫ 1 THz | | | 460(10) GHz | | | this work |
|  | 1486/1633 GHz | | | 1510/1333 GHz | | | 658/1799 GHz | | | CF/*ab-initio*[31] |

In addition to the important role of the transverse components of the anisotropic *g*-tensor for the magnetic relaxation mechanisms, one main difference in the electronic structure of the studied complexes is the presence of particularly low-lying excited KDs. While **(1)** and **(3)** exhibit comparable non-axiality of $g_{eff}$, the energy gap between the ground state and the first excited KD ($\Delta_{1\to2}$) is almost three times larger in **(1)**. Considering the predominantly prolate electron distribution of Er$^{III}$, this observation implies that the additional charge on the apical ligand effectively reduces the equatorial nature of the crystal field acting on the magnetic center. The absence of field-induced slow magnetic relaxation in **(3)** but not in **(1)**, hence, indicates that the dominant relaxation process is thermally activated and involves excited energy levels. Considering the effective energy barriers measured for **(1)** and **(2)**, it seems plausible to attribute fast magnetic relaxation to thermally assisted QTM (TA-QTM), which we conclude to be the main driver of the SMM-silent behaviour in **(3)**. In this respect, we also note that the energy gap between KD1 and KD2 in **(2)** is slightly smaller than in **(1)** and the non-axiality of $g_{eff}$ is more pronounced in **(1)**. In both complexes, KD3 is strongly gapped $\Delta_{1\to3} \gg 1$ THz and likely does not contribute to TA-QTM. While a comparison of $\Delta_{1\to2}$ implies stronger effects of thermally assisted QTM in **(2)** than in **(1)**, the observed smaller $U_{eff}$ in **(1)** highlights the relevance of non-axiality.[31]

For KD1, our experimental results of both the *g*-tensor and the energy gap $\Delta_{1\to2}$ are in reasonably good quantitative agreement with the findings of CF analysis of the dc magnetisation[31], which indicates the validity of the numerical approach for the ground state KD (see Tab. 1). Reported *ab-initio* results for KD1 fairly agree to the actual *g*-values and only yield a rather qualitative prognosis on the magnetic anisotropy in the complexes under study. In particular, the considerable differences between the predictions of *ab-initio* calculations and experimentally obtained parameters for the anionic complex **(3)** highlight the difficulty of including the effects of the total molecular charge on the crystal packing in the quantum chemical calculations of pseudo-isolated complexes.[26, 31] In contrast to the observations for KD1, the parameters of the excited states seem to be better reproduced by *ab-initio* results, while the results of CF analysis, except for a reasonable agreement of the energy differences between the KDs, are quite off from the actually observed values. The observation of CF analysis performing worse for higher-lying KDs is not unsurprising since the contribution of the excited states to the dc magnetisation decreases with their energy difference to the magnetic ground state.

**Table 2.** Effective magnetic quantum numbers $m_{eff}$ for the lowest KDs as obtained from the HF-EPR data assuming $g_L = 1.2$ (see the text).

|  | complex **(1)** | complex **(2)** | complex **(3)** |
|---|---|---|---|
| KD1 | 5.0(1) | 5.5(1) | 5.2(2) |
| KD2 | 2.8(2) | 2.6(1) | 2.0(1) |
| KD3 | - | - | 1.3(1) |

## Conclusion

In summary, we have experimentally investigated the influence of the apical ligand in mononuclear pentagonal-bipyramidal coordinated erbium complexes using HF-EPR. Our spectroscopic data enable us to precisely determine effective $g$-tensors and crystal field splittings of the two energetically lowest Kramers doublets and, hence, allow us to assess the validity of recently reported crystal field analysis and *ab-initio* studies. For all studied compounds, the anisotropic $g$-tensor of the ground state KD was shown to possess high axiality. Yet, substituting the apical ligands from $CH_3OH/N_3$ to $Cl/H_2O$ and $Cl/Cl$ and leaving the equatorial $N_3O_2$ Schiff-Base ligand unchanged introduces transversal anisotropy, thereby leading to a mixing of the *m*-levels. Furthermore, direct observation of the crystal field splittings reveals significantly larger energy separation of the first excited Kramers doublet to the magnetic ground state in the two neutral complexes ($\Delta_{(1)}$ = 825 GHz and $\Delta_{(2)}$ = 767 GHz) as compared to the charged complex ($\Delta_{(3)}$ = 270 GHz). The experimentally found $g$-values suggest clear mixing of states in KD1 and KD2 of complex **(1)**, while half-integer $m_{eff}$ values found for KD1 and KD2 in complex **(2)** indicate almost pure $m$ = 11/2 and $m$ = 5/2 compositions of the magnetic ground state and first excited state, respectively. Combining our findings, the absence of field-induced SMM behaviour in complex **(3)**, but not in **(1)** and **(2)**, as found in ac susceptibility measurements[31], can likely be explained by a comparably small energy barrier along with pronounced transversal anisotropy giving rise to relatively fast relaxation pathways via QTM and/or thermally assisted QTM.

## Experimental Methods

Continuous wave high-frequency/high-field electron paramagnetic resonance (HF-EPR) measurements were performed using a millimetre vector network analyser (MVNA) from *ABmm* as a phase-sensitive microwave source and detector.[35] Thoroughly ground powder samples were prepared inside a brass ring sealed with Kapton tape and placed into the microwave path of cylindrical brass waveguides. The data were obtained by simultaneously measuring the transmitted microwave amplitude and phase at frequencies ranging from 80 GHz to 850 GHz in magnetic fields up to 16 T. All spectra were corrected for phase-mixing and background effects, such as temperature drifts and contributions from the experimental setup.[39] Temperature control down to 2 K was achieved in a variable temperature insert (VTI) of an Oxford magnet system. Powder samples were firstly prepared inside a brass ring without glue or grease, i.e., a loose powder experiment was performed to allow for the alignment of the crystallites along their effective anisotropy axis in the external magnetic field. This technique has proven to strongly reduce the complexity of the so-obtained pseudo-single-crystal spectra, especially for 3$d$ metal-containing compounds.[40-45] Alignment was ensured by applying a magnetic field of 16 T prior to the measurements and restricting the field range to 0.2 - 16 T. To investigate the magnetic anisotropy in all directions of the molecules, further HF-EPR experiments were carried out on powder samples, which were fixed in their initial randomly distributed orientation using n-eicosane ($C_{20}H_{42}$, CAS-Nr. 112-95-8). Spectral simulations were performed using the MATLAB software package EasySpin.[46]

## Acknowledgements

Measurements were performed before February 2022 supported by Deutsche Forschungsgemeinschaft (DFG) under Germany's Excellence Strategy EXC2181/1-390900948 (the Heidelberg STRUCTURES Excellence Cluster). J.~A. acknowledges support by the International Max-Planck Research School for Quantum Dynamics (IMPRS-QD) Heidelberg. E. B. Y. and T. A. B. thank the Ministry of Science and Higher Education of the Russian Federation for financial support (Government Assignment No. 124013100858-3 and 124020200104-8).

# Supporting Information: The Relevance of Non-axiality and Low-lying Excited States for Slow Magnetic Relaxation in Pentagonal-bipyramidal Erbium(III) Complexes Probed by High-frequency EPR

Jan Arneth,*[a] Lena Spillecke,[a] Changyun Koo,[a],[c] Tamara A. Bazhenova,[b] Eduard B. Yagubskii,[b] and Rüdiger Klingeler*[a]

Corresponding authors: jan.arneth@kip.uni-heidelberg.de, ruediger.klingeler@kip.uni-heidelberg.de

**Table of Contents**

## 1. Additional HF-EPR Data on Complex (1)

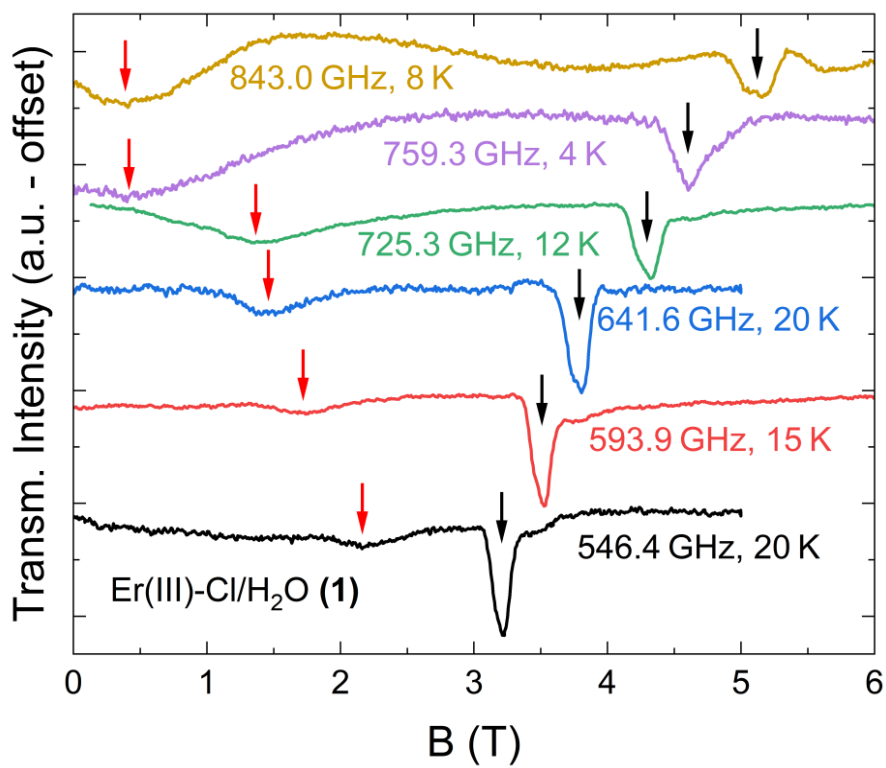

**Figure S1.** Loose powder HF-EPR spectra of complex **(1)** at selected fixed frequencies for $T > 2$ K. Arrows mark the read off resonance positions for the transitions arising from the ground state (black) and from excited states (red) as described in the main text.

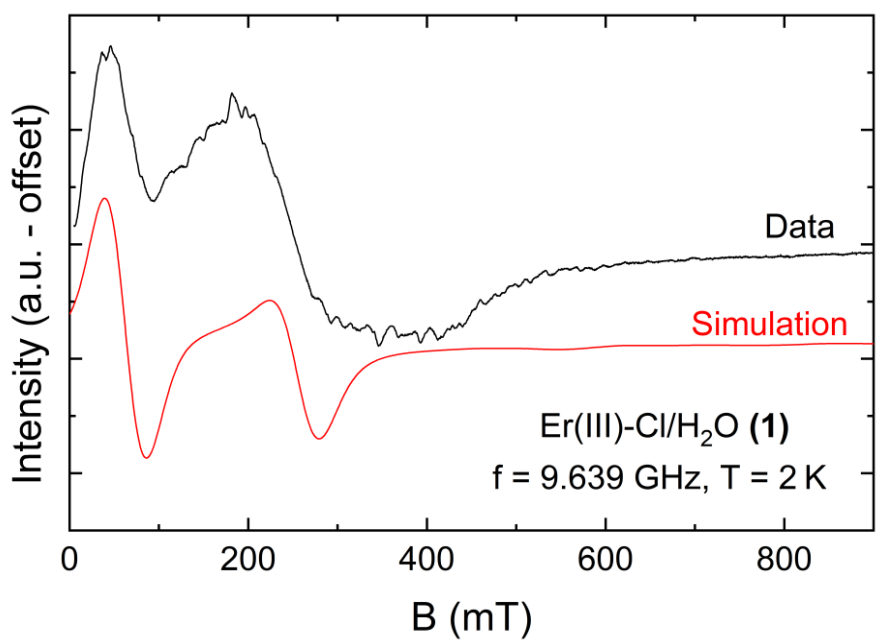

**Figure S2.** Fixed powder X-band spectrum of complex **(1)** at $f = 9.639$ GHz and $T = 2$ K and the corresponding simulation using a $S = 1/2$ pseudospin approximation with the parameters as described in the main text.

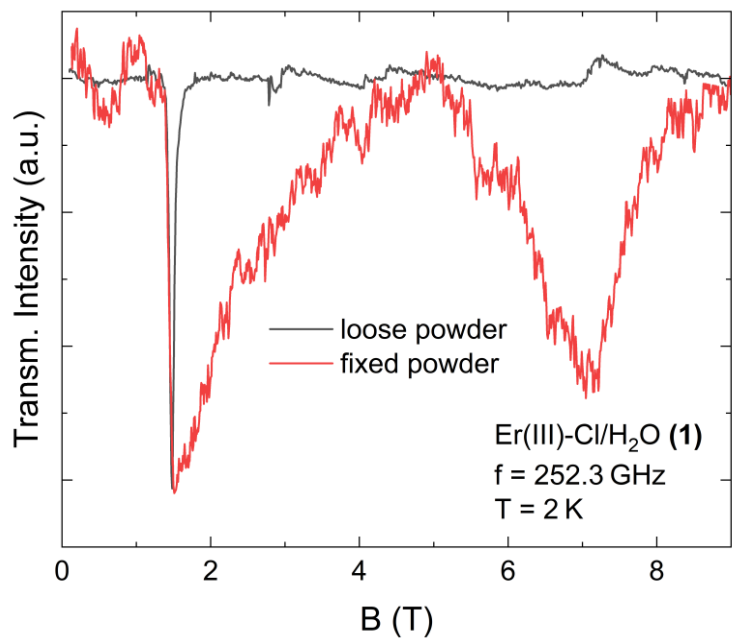

**Figure S3.** Comparison of loose and fixed powder HF-EPR spectra obtained at $f$ = 252.3 GHz and $T$ = 2 K for complex **(1)**. The fixed powder data has been scaled to match the maximum absorption with that of the loose powder data.

## 2. Additional HF-EPR Data on Complex (2)

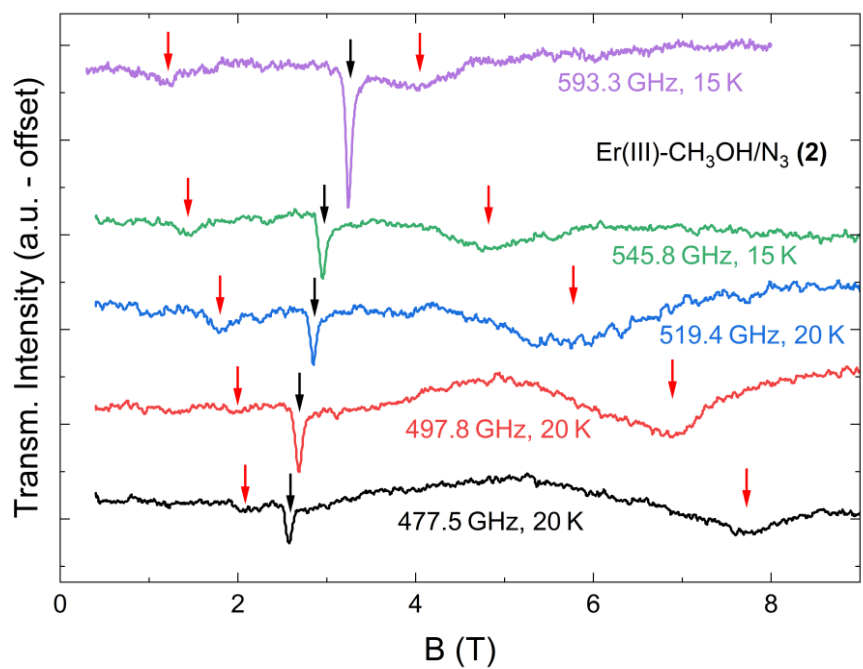

**Figure S4.** Loose powder HF-EPR spectra of complex **(2)** at selected fixed frequencies for $T$ > 2 K. Arrows mark the read off resonance positions for the transitions arising from the ground state (black) and from excited states (red) as described in the main text.

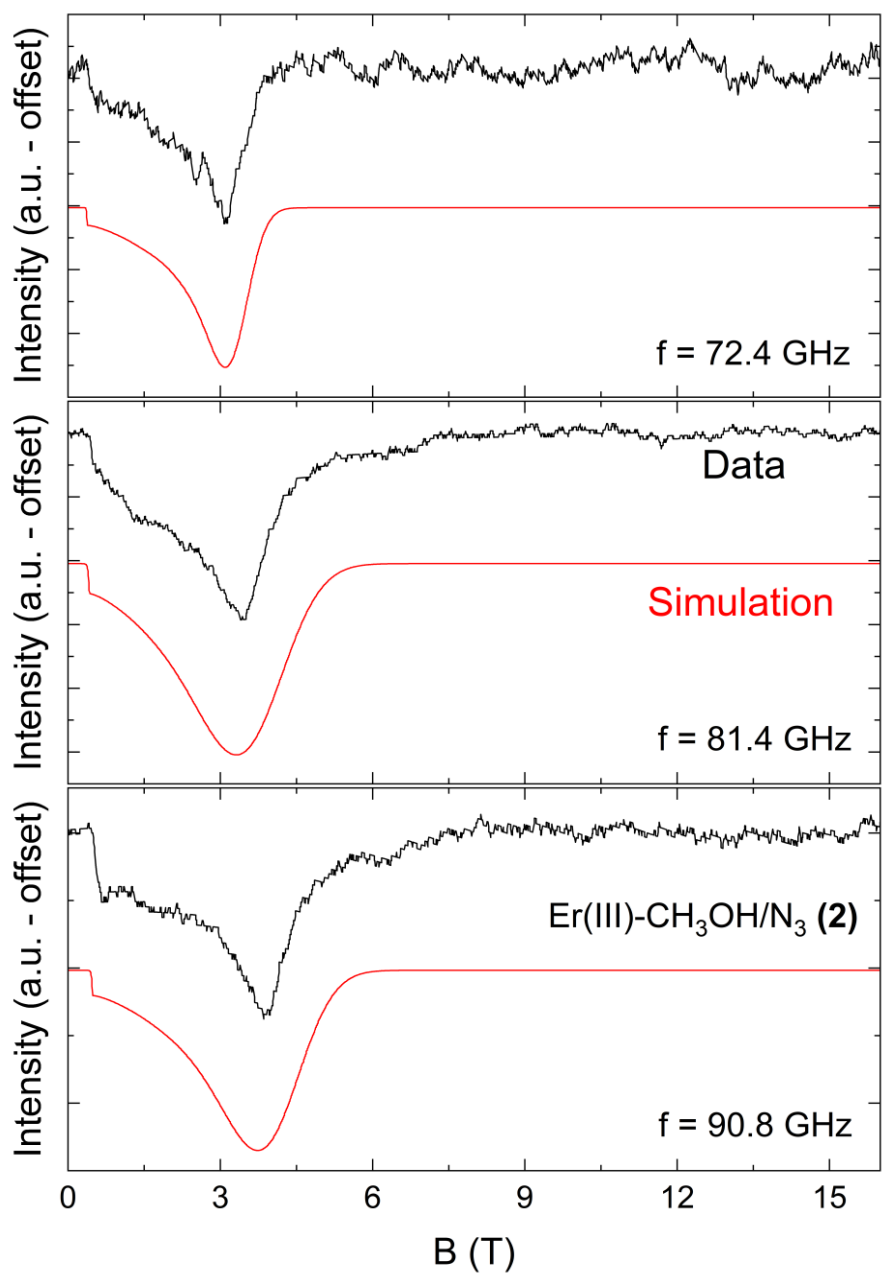

**Figure S5.** Fixed powder HF-EPR spectra of complex **(2)** at selected frequencies and *T* = 2K and simulations using a *S* = 1/2 pseudospin approximation with the parameters as described in the main text.

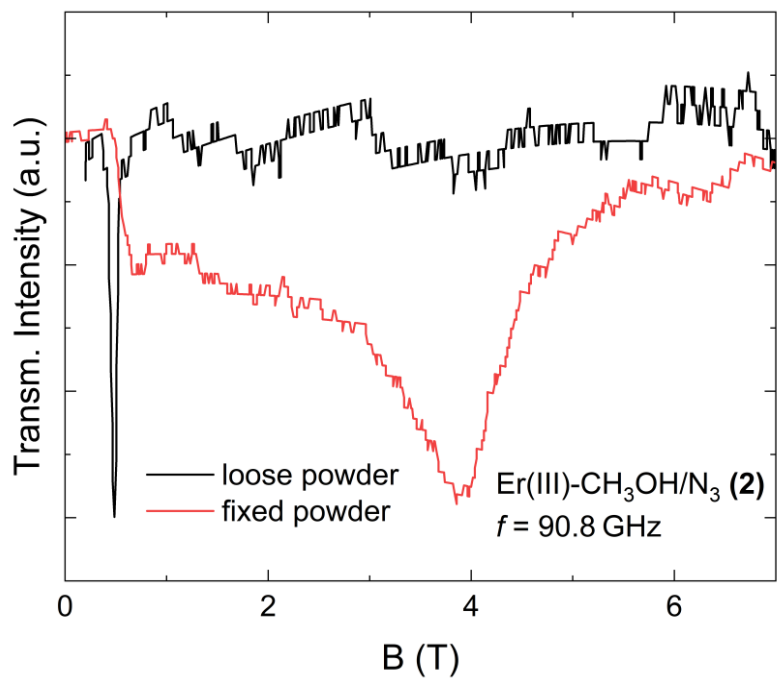

**Figure S6.** Comparison of loose and fixed powder HF-EPR spectra obtained at $f$ = 90.8 GHz and $T$ = 2 K for complex **(2)**. The fixed powder data has been scaled to match the maximum absorption with that of the loose powder data.